\documentclass[10pt,journal]{IEEEtran}

\usepackage{amssymb}
\usepackage{amsmath}
\usepackage{cite}
\usepackage{url}
\usepackage{xcolor}
\usepackage{cite,graphicx,amsmath,amssymb}
\usepackage{fancyhdr}
\usepackage{mdwmath}
\usepackage{mdwtab}
\usepackage{caption}
\usepackage{amsthm}
\usepackage{setspace}
\usepackage{hyperref}
\usepackage{algorithm}
\usepackage{algorithmic}
\usepackage{multirow}
\usepackage{makecell}
\usepackage{mathtools}
\usepackage{subcaption}
\hypersetup{colorlinks=false}

\graphicspath{{./src/img/}}

\newtheorem{theorem}{Theorem}

\newtheorem{lemma}{Lemma}

\newtheorem{corollary}{Corollary}

\allowdisplaybreaks
\title{Terahertz Near-Field Communications and Sensing}

\author{
        Zhaolin~Wang,~\IEEEmembership{Graduate Student Member,~IEEE,}
        Xidong~Mu,~\IEEEmembership{Member,~IEEE,} \\
        and Yuanwei~Liu,~\IEEEmembership{Senior Member,~IEEE}
\thanks{The authors are with the School of Electronic Engineering
and Computer Science, Queen Mary University of London, London E1 4NS, U.K. (e-mail: zhaolin.wang@qmul.ac.uk, xidong.mu@qmul.ac.uk, yuanwei.liu@qmul.ac.uk).}
}

\begin{document}

\maketitle

\begin{abstract}

    This article focuses on the near-field effect in terahertz (THz) communications and sensing systems. 
    By equipping with extremely large-scale antenna arrays (ELAAs), the near-field region in THz systems can be possibly extended to hundreds of meters in proximity to THz transceivers, which necessitates the consideration of near-field effect in the THz band both for the communications and sensing. We first review the main characteristics of the near-field region in the THz bands. The signal propagation in the near-field region is characterized by spherical waves rather than planar waves in the far-field region. This distinction introduces a new distance dimension to the communication and sensing channels, which brings new opportunities and challenges for both THz communications and sensing. More particularly, 1) For THz communications, the near-field effect enables a new mechanism for beamforming, namely, beamfocusing, in the focusing region. Furthermore, in THz multiple-input and multiple-output (MIMO) systems, the near-field effect can be exploited to combat the multiplexing gain degradation caused by the sparse THz channels. To address the near-field beam split effect caused by the conventional frequency-independent hybrid beamforming architecture in THz wideband communications, we propose a pair of wideband beamforming optimization approaches by a new hybrid beamforming architecture based on true-time-delayers (TTDs). 2) For THz sensing, joint angle and distance sensing can be achieved in the near-field region. Additionally, the near-field beam split becomes a beneficial effect for enhancing the sensing performance by focusing on multiple possible target locations rather than a drawback encountered in communications. Finally, several topics for future research are discussed.

\end{abstract}

\section{Introduction}
Wireless networks are witnessing the emergence of various application scenarios, including augmented reality, remote healthcare, smart cities, autonomous vehicles, and industrial automation. In addition to essential communications, sensing is poised to play a crucial role in enabling and enhancing these applications, which motivates the heated discussion of integrated sensing and communications (ISAC) recently \cite{huawei2022}. On the one hand, ISAC allows for dynamic spectrum sharing and optimization between these two functions, ultimately enhancing spectrum efficiency. On the other hand, ISAC can also facilitate the mutual assistance between these two functions, and thereby achieve the coordination gain. For example, the next-generation networks are expected to support a wide range of applications with diverse quality of service (QoS) requirements, including ultra-reliable and low-latency communications (URLLC). By integrating sensing capabilities, the wireless network can continuously monitor and adapt to changing environmental conditions, interference levels, and user demands, enabling proactive resource allocation and optimization to meet stringent QoS requirements. Furthermore, by fusing multiple sources of data within well-established communication networks, such as signals from base stations, user equipment, and other devices, more precise and robust sensing can be achieved. 

\begin{figure*}[t!]
    \centering
    \includegraphics[width=0.7\textwidth]{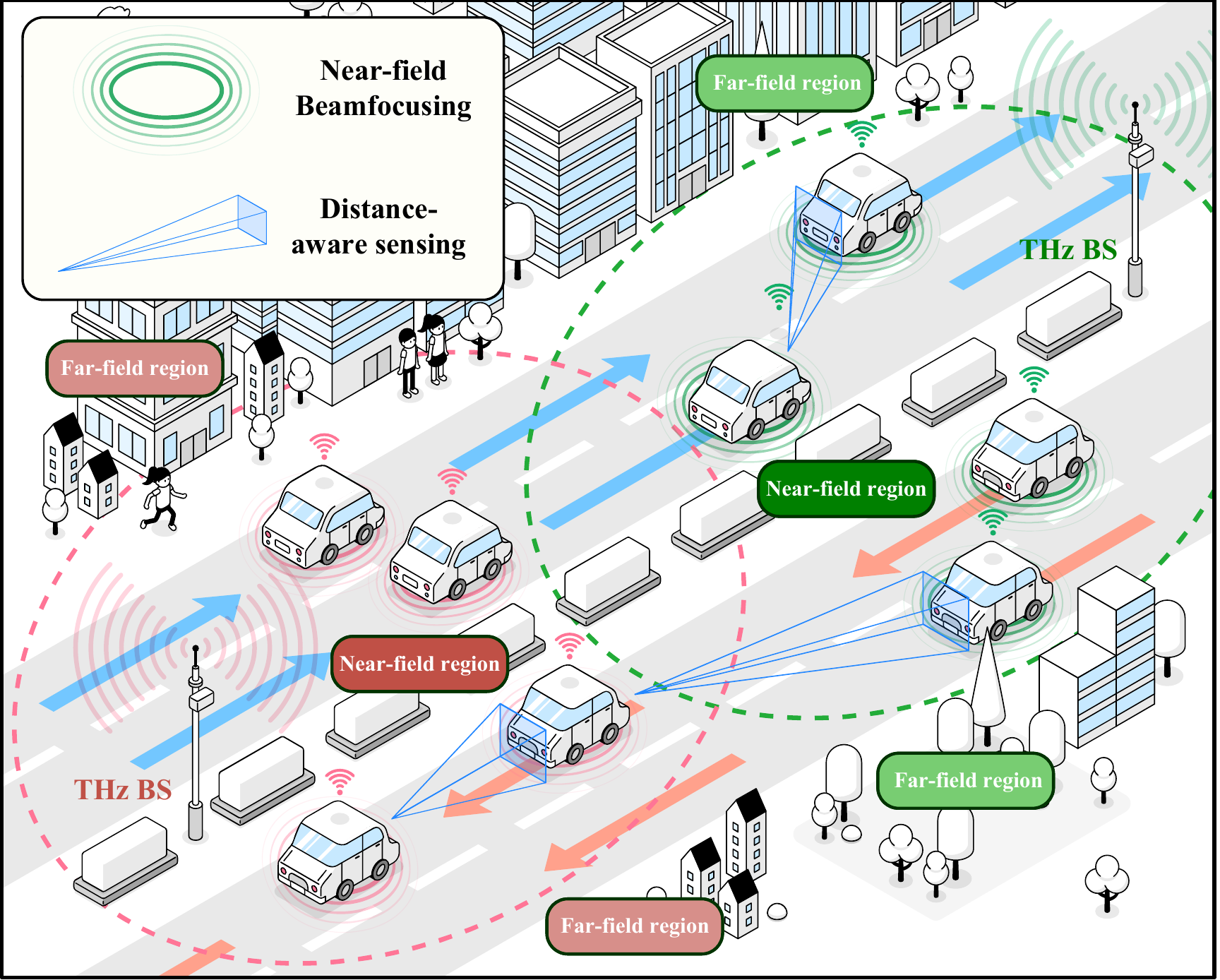}
    \caption{Illustration of the THz near-field communications and sensing. The near-field region in THz systems can be in the order of hundreds of meters.}
    \label{fig:system}
\end{figure*}

To meet the rigorous demands of communications and sensing in the next-generation wireless networks, a large bandwidth is required to boost the communication capacity and sensing resolution. The ultra-wide terahertz (THz) band spanning from $0.1$ THz to $10$ THz, offers vast bandwidth resources in the order of tens of gigahertz (GHz) for communications \cite{yang2022terahertz}. Furthermore, THz waves can provide optics-like ultra-high sensing resolution but with good penetration ability for opaque materials \cite{huawei2022}. Hence, the THz band has become a promising frequency band for communications and sensing. While there are significant benefits to using the THz band, challenges remain, such as propagation losses over long distances, bandwidth variations dependent on the range, and the need for compact hardware implementation. In particular, to combat the severe propagation loss in the THz band, the extremely large-scale antenna arrays (ELAAs), which can consist of hundreds of or even thousands of antenna elements, are typically exploited to generate pencil-like beams and enhance array gain. However, the utilization of ELAAs in the THz band introduces significant near-field effects compared to conventional frequency bands. Typically, Rayleigh distance is employed as a phase-error-based boundary to differentiate between the near-field and far-field regions, which is increased with the aperture of the antenna array and the operating frequency \cite{zhang20236g}. As a result, the near-field region can be significantly large in THz systems. For instance, consider a ELAA with an aperture of $0.5$ m operating at $0.3$ THz. The Rayleigh distance of this ELAA is around $500$ m. Based on the fact the transmission range of THz systems is generally limited to tens of meters \cite{yang2022terahertz}, the electromagnetic field over the THz band is mainly characterized by the near-field region. However, the propagation characteristics of electromagnetic waves in the near-field region are fundamentally different from those in the far-field region. Consequently, the design of communications and sensing in the THz band need to be reconsidered from the near-field perspective. Furthermore, the near-field effect in THz systems opens new opportunities for both communications and sensing, such as beamfocusing and distance-aware sensing shown in Fig. \ref{fig:system}, as well as several challenges, which will be detailed in the following sections.

In this article, we present a comprehensive investigation into the influence of near-field effects on both communications and sensing in the THz band. We begin by providing a systematic review of the fundamental characteristics of the near-field. Building upon these characteristics, we then highlight the primary opportunities and challenges that arise in the context of THz near-field communications and sensing. Finally, the topics for future research are discussed.

\section{THz-Band Characteristics: From Far-Field to Near-Field}
Although the THz band is capable of providing huge bandwidth resources for communications and sensing, its near-field characteristics present additional challenges in system design that are not encountered in lower frequencies. Understanding the fundamental near-field characteristics of THz channels is critical for communication and sensing systems design in the THz band. 

\begin{figure*}[t!]
    \centering
    \includegraphics[width=1\textwidth]{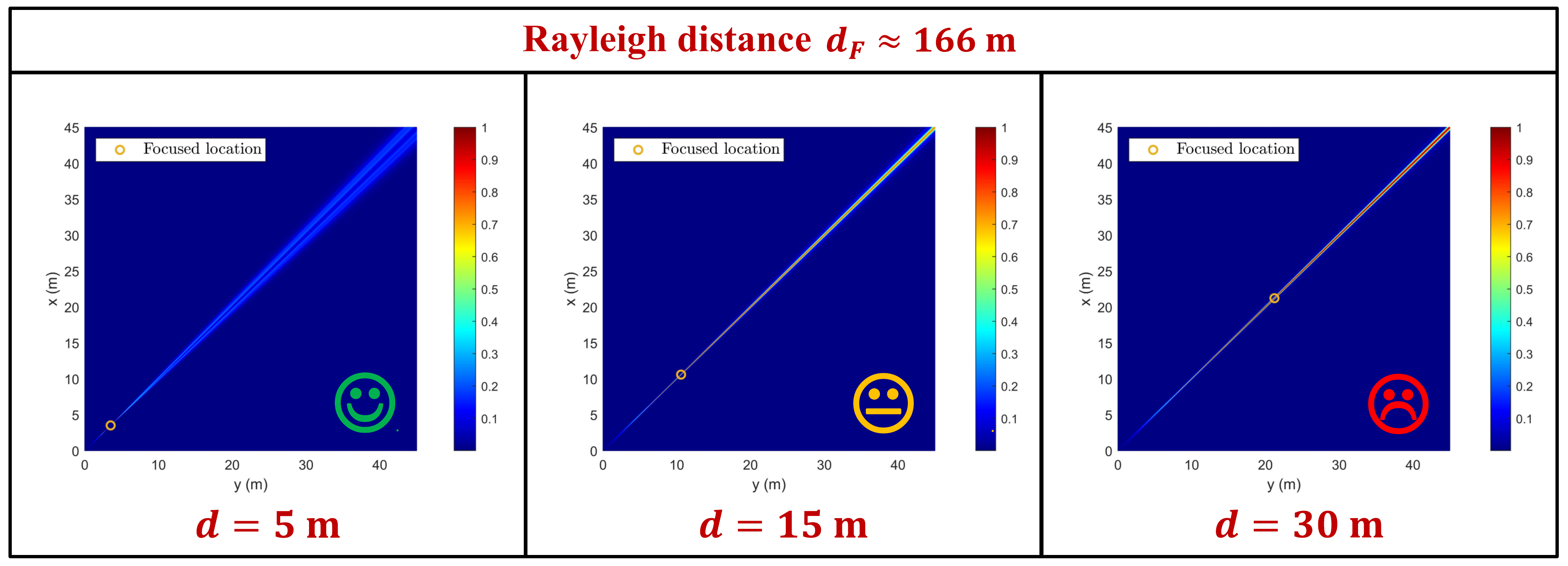}
    \caption{Performance of THz near-field beamfocusing at different distances. The simulation
    setup is given as follows. The transmitter is equipped with a uniform linear array with an aperture of $0.5$ m and half-wavelength antenna spacing. The operating frequency is $0.1$ THz. }
    \label{fig:focusing_region}
\end{figure*}

\subsection{Spherical Wave Propagation}
As previously discussed, the primary difference between the near-field and far-field regions lies in the manner of signal propagation. In particular, in the near-field region, signal propagation is based on \emph{spherical waves}, while in the far-field region, it relies on \emph{planar waves}. Nonetheless, it is important to highlight that the spherical wave serves as the ground truth in both regions, while the planar wave is only a practical approximation of the spherical wave when the propagation distance is significantly large. More specifically, the far-field planar-wave model is obtained by approximating the channel phase in the near-field spherical-wave model through the first-order Taylor expansion. The Rayleigh distance is defined as the minimum distance such that the phase approximation error does not exceed $\pi/8$. Consequently, employing the far-field planar-wave model in the THz system design can lead to notable performance degradation due to its inconsistency with reality. A recent study has demonstrated that this inconsistency can result in a substantial underestimation of the capacity performance in systems operating at frequencies of $30$ GHz or above, even when operating beyond the Rayleigh distance \cite{10104065}. Therefore, near-field spherical wave propagation becomes a must-considered factor in the system design of THz communications and sensing.

\subsection{New Distance Dimension}
Although the near-field spherical wave model is more complex than the far-field planar wave model and causes new challenges, it opens new opportunities for designing communications and sensing in the THz band. In particular, for far-field planar wave propagation, the signal has a \emph{linear phase} with respect to the antenna index. Thus, the far-field channel is characterized by the \emph{angle} between the transmitter and receiver. On the contrary, the spherical wave propagation in the near-field region leads to a \emph{non-linear phase} of the signal with respect to the antenna index. As a result, apart from the angle dimension, there is a new \emph{distance} dimension in near-field channels. The new distance dimension is capable of enabling distance-aware communications and sensing in the THz bands, which will be further discussed in the following.

\section{THz Near-Field Communications}
In this section, we discuss the potential of the near-field effect in enhancing the connectivity and multiplexing gain of THz communications. Furthermore, regarding the undesired near-field beam split effect in THz wideband communications, two novel wideband near-field beamforming approaches are proposed by exploiting true-time delayers (TTDs).

\subsection{Connectivity Enhancement by Beamfocusing}
THz communication is envisioned to support a significant number of users in particular scenarios, such as large gatherings and intelligent factories with massive connected mobile devices or machine-type devices, by exploiting the abundant spectrum resources at the THz band. To this end, various efficient spectrum allocation schemes have been proposed for THz communications, including the utilization of multiple transmission windows (TWs) and frequency reuse \cite{yang2022terahertz}. In particular, the unique frequency-selective THz molecular absorption loss divides the THz band into several TWs. Within each TWs, multiple users can be served by multiple non-overlapping sub-band with high data rates. However, as the number of user equipment increases and communication requirements becomes more stringent, it becomes imperative to implement frequency reuse among users within THz TWs. In this case, the inter-user interference needs to be effectively mitigated to avoid significant performance loss. 

Thanks to the ELAAs exploited in THz communications, non-overlapping narrow beams can be generated to distinguish users in the spatial domain and facilitate frequency reuse \cite{yang2022terahertz}. However, in conventional far-field THz communications, the generated beams can only steer towards a specific direction due to the planar wave propagation, which is referred to as far-field \emph{beamsteering}. Consequently, users located in a similar direction may experience significant interference in the spatial domain, necessitating the utilization of more sophisticated multiple access techniques like non-orthogonal multiple access (NOMA) distinguishing users in the power domain or code domain. Fortunately, the propagation of spherical waves induced by the near-field effect introduces an additional distance dimension to wireless channels. This new dimension enables distance-aware beamforming in near-field THz communications, which is fundamentally different from far-field beamforming. Specifically, near-field beamforming has the capability to focus on a specific location, known as near-field “beamfocusing” \cite{zhang2022beam}. Consequently, users can be distinguished in the polar domain, encompassing both distance and angle information, providing greater flexibility for mitigating inter-user interference. Therefore, near-field beamfocusing emerges as a promising technique to facilitate extensive connectivity in THz communications.

However, it should be noted that achieving near-field beam focusing is limited to a specific region within the near-field, rather than being applicable throughout the entire region. As shown in Fig. \ref{fig:focusing_region}, despite the Rayleigh distance potentially spanning hundreds of meters in THz systems, the “focusing” effect only occurs within a range of tens of meters. Consequently, in order to leverage near-field beamfocusing in THz communications, the antenna array must be exceptionally large to generate a sufficiently sizable “focusing region”. Furthermore, to effectively support users located in a similar direction outside the “focusing region”, NOMA techniques can be invoked. In this case, the complexity of the beamforming design through the conventional optimization technique may be unacceptable. Therefore, the development of low-complexity beamforming designs is crucial to facilitate near-field beamfocusing in THz communications. 

\begin{figure*}[t!]
    \centering
    \includegraphics[width=1\textwidth]{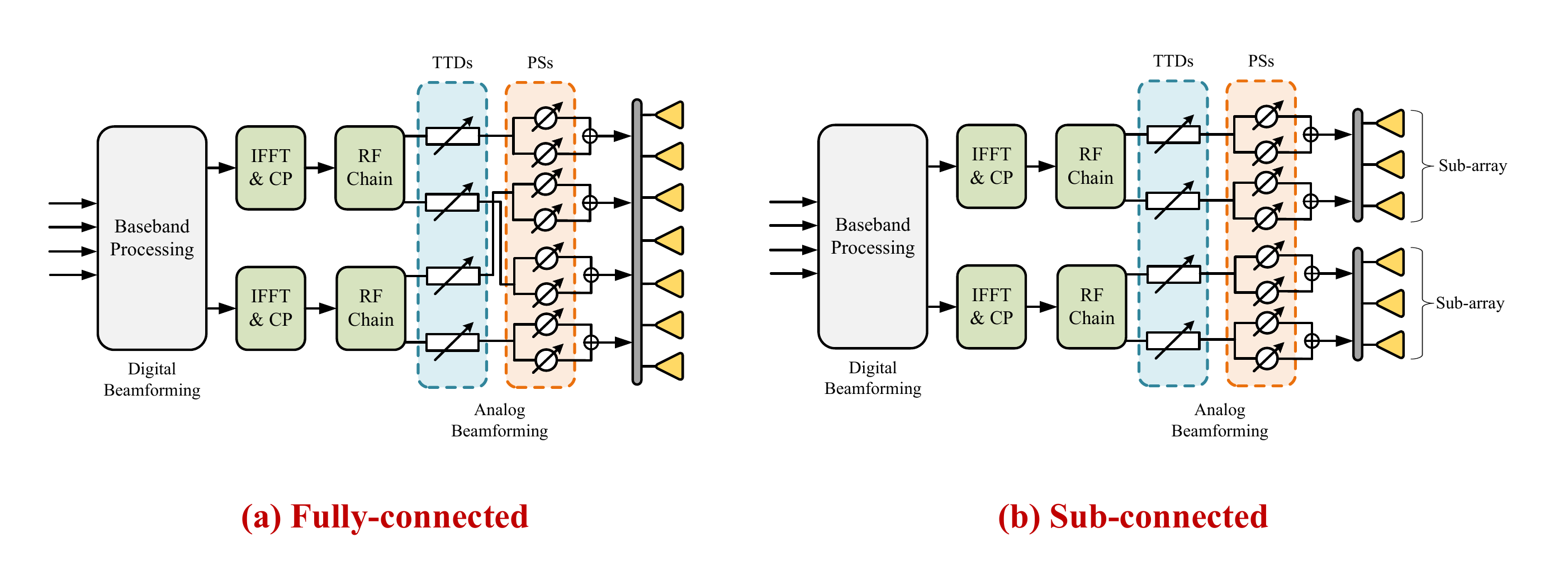}
    \caption{TTD-based hybrid beamforming architectures for THz wideband communications. (a) Fully-connected architecture: each RF chain is connected to the entire antenna array via PSs and TTDs. (b) Sub-connected architecture: each RF chain is connected to a sub-array via PSs and TTDs.}
    \label{fig:architecture}
\end{figure*}

\subsection{Multiplexing Gain Enhancement in THz MIMO Systems}

Multiple-input multiple-output (MIMO), which exploits multiple antennas at both transmitter and receiver for achieving enhanced multiplexing gain in wireless communications, encounters certain performance challenges when applied to THz communications that rely on the far-field model. In THz communications, the presence of significant scattering, diffraction, and reflection losses results in weakened non-line-of-sight (NLoS) channels \cite{yang2022terahertz}. Consequently, the THz communication channel has a sparse nature, with the line-of-sight (LoS) channel dominating the overall link characteristics. For conventional far-field channel models, the LoS channel matrix is inherently rank-one due to the linear phase caused by planar wave propagation. As a consequence, despite the existence of multiple antennas at both the transmitting and receiving ends, the system can support only a single data stream, leading to a considerably diminished multiplexing gain. This constraint restricts the complete utilization of the advantages of MIMO in THz communications based on the far-field model. Fortunately, in practical THz communications scenarios, the near-field model becomes more applicable. The non-linear phase resulting from near-field spherical wave propagation introduces changes to the LoS channel matrix, making it no longer rank-one. Consequently, the near-field LoS channel becomes capable of supporting multiple data streams without the aid of NLoS channels, thereby enhancing the achievable multiplexing gain. It is worth noting that the sparsity of channels in THz communications is no longer a substantial obstacle in this context. While the near-field LoS channel has a high rank, its rank gradually diminishes as the distance increases. This fact introduces new challenges for the design of hybrid digital and analog beamforming, where the number of RF chains should be adjusted adaptively according to the distance to achieve the optimal multiplexing gain without energy waste.

\subsection{Near-Field Beam Split in THz Wideband Systems}

One of the primary motivations for exploiting the THz band is its ability to provide abundant bandwidth resources. Hence, it becomes imperative to consider the wideband effect during the design of THz systems.
In addition to the frequency-selective wideband effect, significant propagation delays across the ELAAs lead to the spatial-selective wideband effect. In simpler terms, a user's spatial position in relation to the base station varies across different frequencies. Consequently, when utilizing the popular hybrid digital and analog beamforming architecture, the beams produced at different frequencies may not perfectly align with the user's location, as the frequency-independent phase shifters (PSs) are exploited to generate the same analog beamformer for all frequencies. The spatial selectivity mentioned above is generally not a significant concern for traditional frequency bands like the millimeter wave (mmWave) band, as the main lobes of the beams at different frequencies can still encompass the user's location. However, in THz systems, the combination of an exceptionally wide bandwidth and antenna array leads to the occurrence of the \emph{beam split} effect \cite{dai2022delay}, where the main lobes of the beams at the majority of frequencies fail to cover the user's location, resulting in undesirable array gain loss. Moreover, it is important to note that the beam split effect only takes place in the angular domain within the far-field region but also extends to the distance domain within the near-field region. Therefore, the near-field beam split effect is much more aggregated and challenging to solve.

\begin{figure}[t!]
    \centering
    \begin{subfigure}[t]{0.4\textwidth}
        \centering
        \includegraphics[width=1\textwidth]{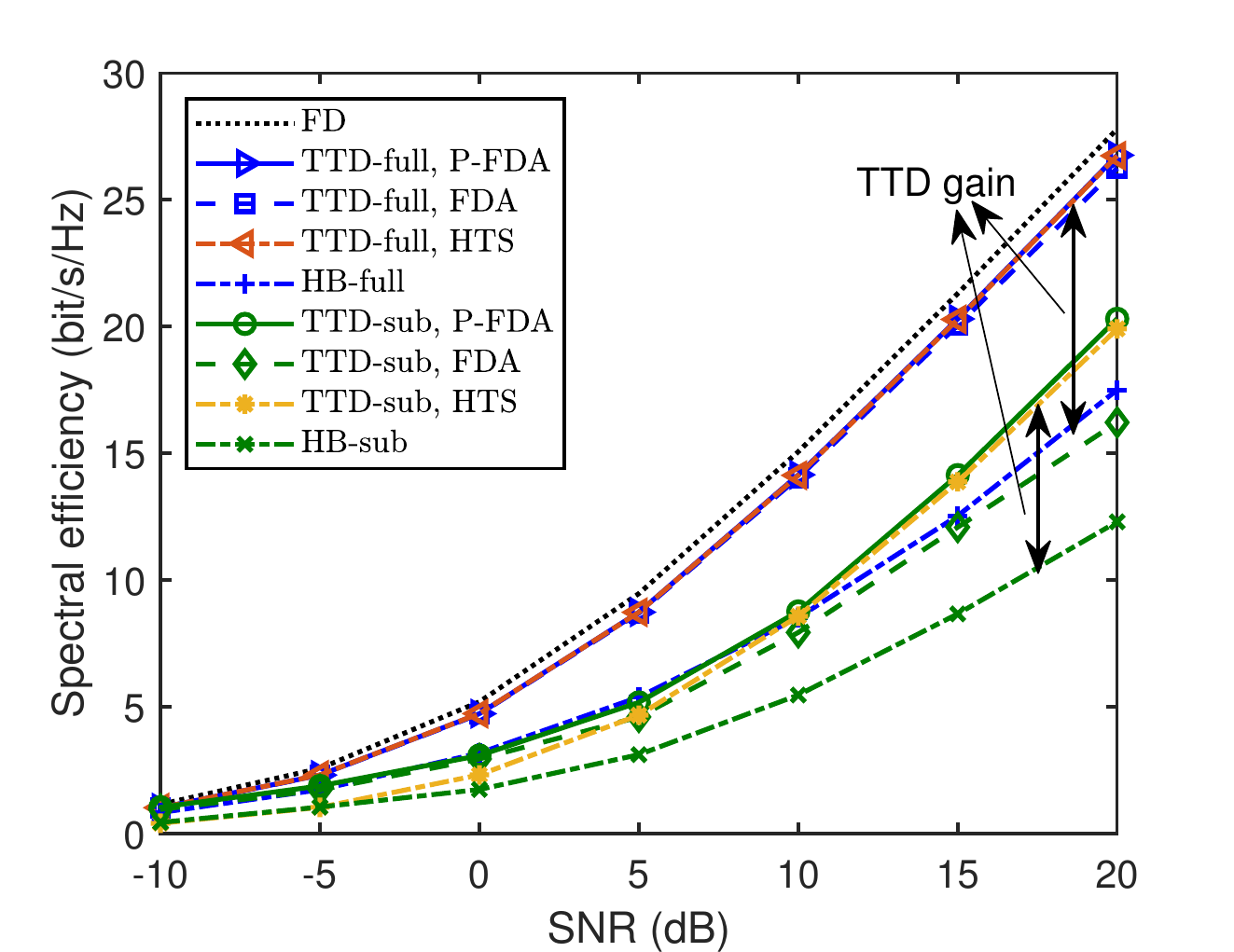}
        \caption{Spectral efficiency versus SNR}
        \label{fig:TTD_SNR}
    \end{subfigure}
    \begin{subfigure}[t]{0.4\textwidth}
        \centering
        \includegraphics[width=1\textwidth]{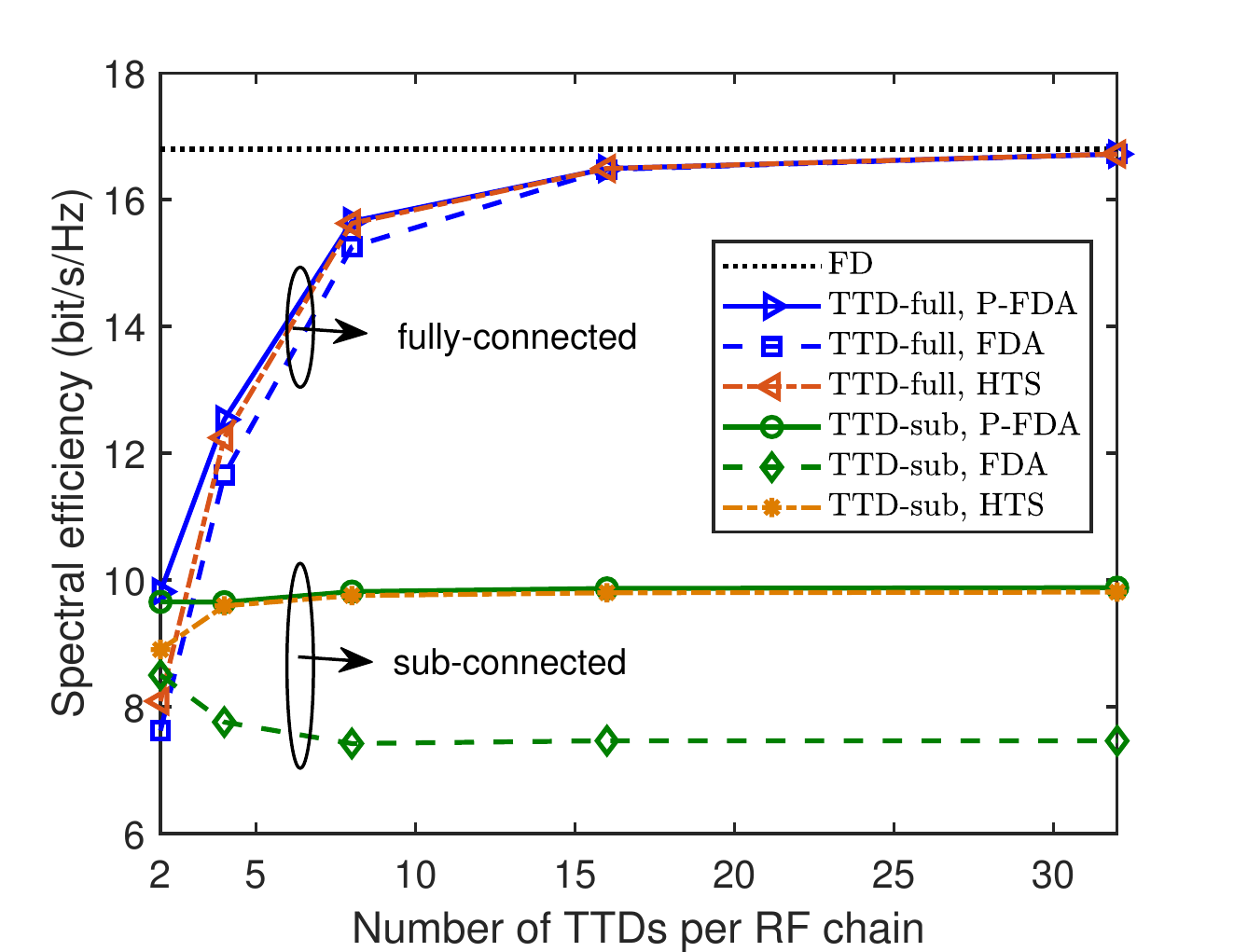}
        \caption{Spectral efficiency versus the number of TTDs per RF chain.}
        \label{fig:TTD_number}
    \end{subfigure}
    \caption{Performance of TTD-based hybrid beamforming architectures in the THz near-field wideband systems. Specifically, “FD” represents the optimal fully-digital beamforming. “TTD-full” and “TTD-sub” represent the fully-connected and sub-connected TTD-based hybrid beamforming, respectively. “HB-full” and “HB-sub” represent the fully-connected and sub-connected conventional hybrid beamforming, respectively. “P-FDA” represents the penalty-based FDA approach.}
    \label{fig:TTD}
\end{figure}

Several methods have been proposed for mitigating the beam split effect, the key idea of which is to realize the frequency-dependent analog beamformer. One of the most popular methods is to exploit the TTDs in the analog beamformers \cite{rotman2016true}. Unlike PSs, TTDs have the ability to introduce time delays to the signals, which subsequently become frequency-dependent phase shifts in the frequency domain. However, the power consumption of TTDs is much higher than the PSs. Therefore, a TTD-based hybrid beamforming architecture has been proposed recently for THz wideband systems, where a minimum number of TTDs is inserted between the RF chains and PSs in the conventional hybrid beamforming architectures to facilitate the frequency-dependent analog beamforming. As illustrated in Fig. \ref{fig:architecture}, the TTD-based hybrid beamforming architecture can be either fully-connected or sub-connected \cite{liu2023near}. Compared to the fully-connected architecture, the sub-connected architecture is more energy efficient by exploiting much fewer PSs and TTDs, which will be detailed in the following.

To effectively mitigate the near-field beam split effect in THz wideband communications, careful design of TTD-based hybrid beamforming is required. In this regard, we propose two approaches: the fully-digital approximation (FDA) approach and the heuristic two-stage (HTS) approach. The FDA approach is to optimize hybrid beamformers to approximate the optimal fully-digital beamformers, which has proven effective in designing conventional hybrid beamformers for narrowband systems \cite{el2014spatially}. By incorporating the penalty method, the FDA approach can even achieve the Karush-Kuhn-Tucker (KKT) optimal solutions \cite{shi2018spectral}. However, the FDA approach requires the optimization of the high-dimensional fully-digital beamformer and analog beamformer as well as the full channel state information, leading to exponential computational complexity and system overhead. Furthermore, the performance of the FDA approach in optimizing TTD-based hybrid beamformers for near-field THz wideband systems remains an open problem. On the other hand, in the HTS approach, the TTD-based analog beamformers are designed in the first stage to generate beams focusing on the users' locations at all subcarriers. It is worth noting that only the location information of the users is required in this stage, which can be obtained by the low-complexity near-field beam training \cite{zhang2022fast}. Then, in the second stage, the low-dimensional equivalent channels, which involve both the analog beamformer and real channels, are estimated and the low-dimensional digital beamformer is optimized based on the equivalent channels. Consequently, the HTS approach can significantly reduce both the optimization complexity and channel estimation complexity.

Fig. \ref{fig:TTD} demonstrates the performance of TTD-based hybrid beamforming architectures in the near-field THz wideband systems. As shown in Fig. \ref{fig:TTD_SNR}, the TTD-based hybrid architectures achieve superior performance gain compared to conventional hybrid beamforming architectures, thanks to the frequency-dependent analog beamformers implemented by TTDs. While the sub-connected architecture leads to a significant decrease in spectral efficiency compared to the fully-connected architecture, it is important to highlight its advantage in terms of lower power consumption, resulting in higher energy efficiency. Additionally, the penalty-based FDA approach always has the best performance, as it can obtain the KKT optimal solution. Although the analog beamformer is not directly optimized, the HTS approach can still achieve a close performance to the penalty-based FDA approach. It is interesting to observe that without the aid of the penalty method, the FDA approach has the worst performance, especially for the sub-connected architecture. This result implies that the FDA approach in wideband systems is not as efficient as that in narrowband systems.
To obtain more insights, we study the impact of the number of TTDs connected to each RF chain in Fig. \ref{fig:TTD_number}. As can be observed, leveraging additional TTDs improves the performance of the fully-connected architecture, but does not have the same effect on the sub-connected architecture. This is because, for the sub-connected architecture, the TTDs connected to each RF chain only need to combat the near-field beam split effect of a small sub-array, which is considerably less significant compared to the effect exhibited by the entire antenna array. As a result, only a minimal number of TTDs are required to for the sub-connected architecture, which saves more power.

\section{THz Near-Field Sensing}
In this section, the benefit of the additional distance dimension introduced by the near-field effect to THz sensing is highlighted. Additionally, the potential of the near-field beam split effect in enhancing the performance of THz sensing is discussed. 

\subsection{Joint Angle and Distance Sensing}

In the preceding section, we have demonstrated the potential of spherical wave propagation in near-field THz systems to enhance communication performance. From a sensing perspective, spherical wave propagation also presents novel opportunities. In traditional far-field sensing, since the sensing channel merely involves angle information, the angle and distance of targets are typically determined through angle-of-arrival (AOA) estimation and time-of-arrival (TOA) estimation, respectively. By leveraging the estimated angle and distance information, the approximate location of the targets can be inferred in three-dimensional (3D) space. However, the accurate TOA estimation has stringent requirements on the system bandwidth and clock synchronization. On the contrary, in near-field sensing, the sensing channel also encapsulates the distance information due to spherical wave propagation. This characteristic makes it possible to estimate distance without the estimation of TOA and facilitates the \emph{joint angle and distance estimation}, i.e., \emph{near-field sensing}. Near-field sensing is an old problem, which can date back 30 years. In early research \cite{huang1991near}, the application of multiple signal classification (MUSIC) and maximum likelihood estimation (MLE) algorithms, which had been widely employed in far-field sensing scenarios, was investigated for near-field sensing. Their results demonstrated the feasibility of joint angle and distance estimation in the near-field region. However, due to the limited near-field region in the past wireless system, near-field sensing was not applicable in practice. The ELAAs and high frequencies in the THz band provide new opportunities for near-field sensing, which ensures a large valid region of near-field sensing.

\begin{figure}[t!]
    \centering
    \begin{subfigure}[t]{0.4\textwidth}
        \centering
        \includegraphics[width=1\textwidth]{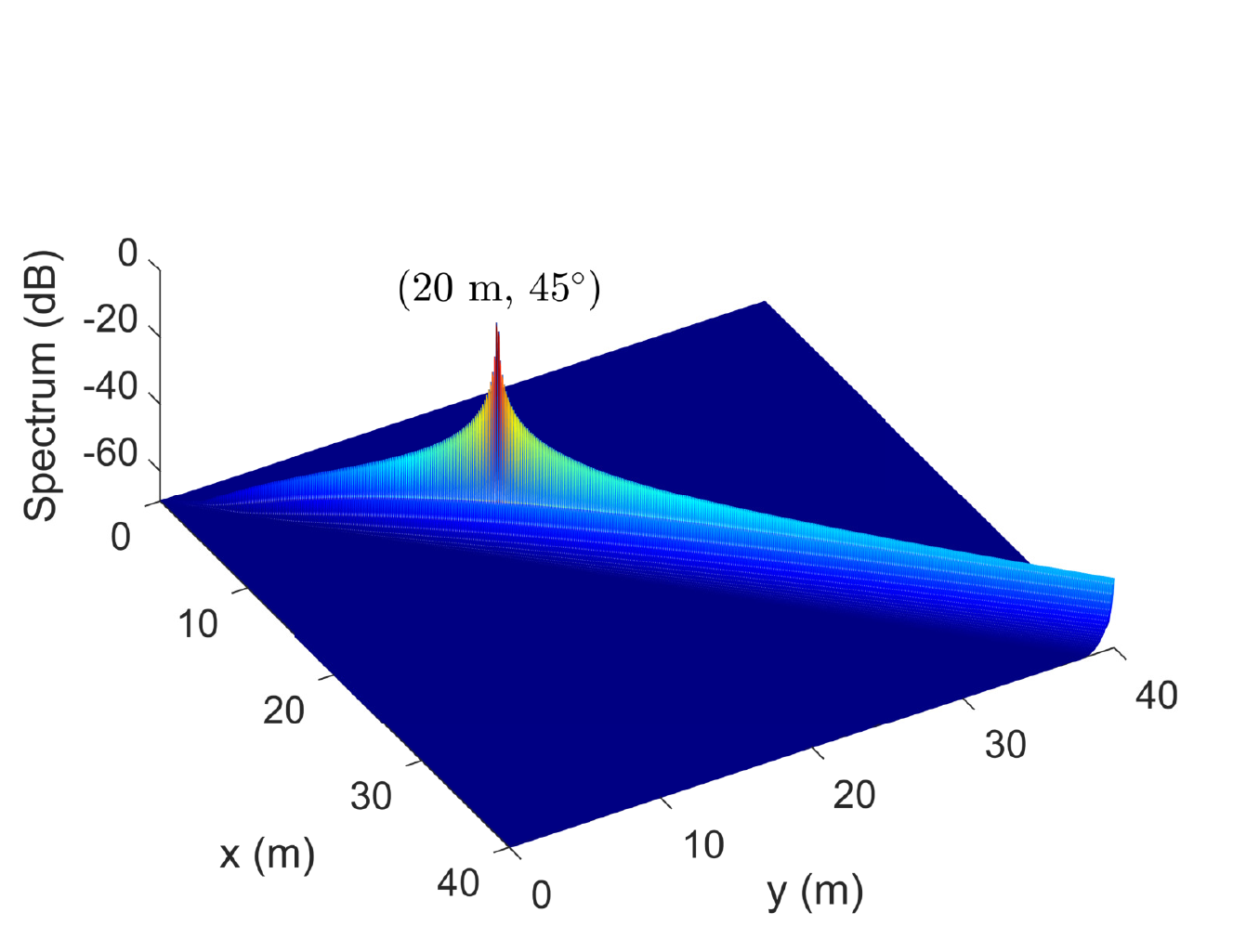}
        \caption{Normalized spectrum of MUSIC.}
        \label{fig:MUSIC}
    \end{subfigure}
    \begin{subfigure}[t]{0.4\textwidth}
        \centering
        \includegraphics[width=1\textwidth]{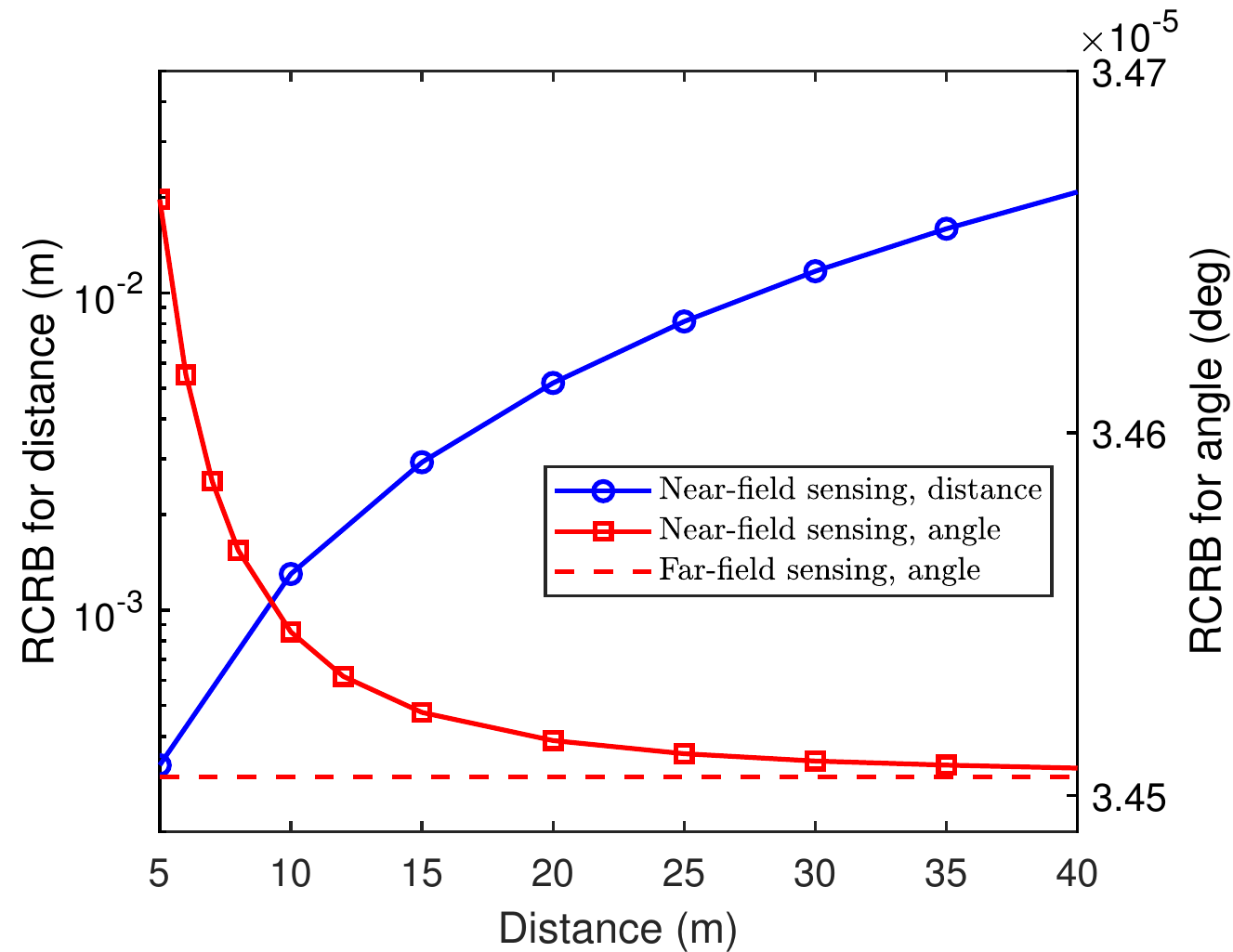}
        \caption{Root Cram{\'e}r-Rao bound (RCRB) versus distance}
        \label{fig:sensing_distance}
    \end{subfigure}
    \caption{Performance of near-field sensing. The simulation setup can be found in \cite{wang2023near}.}
    \label{fig:sensing}
\end{figure}

Fig. \ref{fig:sensing} illustrates the performance of near-field sensing. Specifically, Fig. \ref{fig:MUSIC} presents the estimation result obtained by the MUSIC algorithm in near-field sensing. In this figure, a single target is located at a distance of 20 meters and an angle of 45 degrees away from the sensing node. As can be observed, the spectrum of MUSIC shows a peak response at the target's location, thus confirming the feasibility of near-field sensing for joint angle and distance estimation. This ability makes near-field sensing fundamentally different from far-field sensing, which requires accurate TOA estimation to achieve distance sensing. To obtain more insights, we examine the impact of distance on the performance of near-field sensing in Fig. \ref{fig:sensing_distance}. Here, the Cram{\'e}r-Rao bound (CRB) is exploited as the performance metric, which is a theoretical lower bound of the mean square error (MSE) of parameter estimation by any unbiased estimator. It can be observed that the distance estimation has degraded performance as the distance increases. This is indeed expected because the sensing channel contains less distance information as the target approaches the far-field region. Furthermore, it is interesting to observe that, in terms of angle estimation, the performance of near-filed sensing is bounded by the performance of far-field sensing from below. This is due to the assumption made in far-field sensing, where the incident signals at all antenna elements are assumed to originate from the same direction. However, in near-field sensing, the angle of the incident signal at different antenna elements varies, especially when the target is in close proximity to the sensing node. Consequently, far-field sensing outperforms near-field sensing in terms of angle estimation. But it is worth noting that the performance degradation for angle estimation caused by the shorter distance is merely on the order of $10^{-7}$, which is negligible.

\subsection{Near-Field Beam Split Enhanced THz Sensing}
As discussed earlier, in THz wideband systems, the utilization of the hybrid digital and analog beamforming architecture can lead to the near-field beam split effect. This effect is particularly problematic for communication purposes because the defocusing of beams at certain subcarriers makes them unusable for reliable communication. However, the situation is different for sensing applications, which operate based on a distinct mechanism compared to communication. Sensing aims to gather information from channels, whereas communication aims to transmit information through channels. Consequently, while communication requires stable channels, sensing benefits from “unstable” channels that contain more potential information. From this perspective, the near-field beam split effect can actually enhance the performance of sensing. By generating beams that occupy different locations across various subcarriers, the sensing node has the potential to gather more comprehensive information from the surrounding environment. Furthermore, by exploiting TTDs, the occupied location at different subcarriers can be adjusted adaptively \cite{10058989}, further enhancing sensing flexibility. The advantages of near-field beam split in enhancing sensing capabilities can be illustrated through the following two examples.

\begin{itemize}
    \item \emph{Extended target sensing:} Conventionally, extended target sensing requires the use of a wide beam to encompass the target and resolve its profile. However, in this approach, the echo signals reflected by the resolvable scatterers of the target become mixed, potentially diminishing the overall sensing performance. By leveraging the near-field beam split effect, narrow beams can be generated to focus on distinct locations of the extended target across different subcarriers. Consequently, the echo signals associated with different resolvable scatterers at various subcarriers can be processed independently, thereby significantly enhancing the sensing resolution.
    \item \emph{Moving target sensing:} Moving target sensing is another challenging task for THz sensing. In this task, tracking algorithms are commonly employed to ensure that the sensing beams remain focused on the moving target across different time slots. These algorithms predict the target's location in the next time slot based on the sensing results obtained in the current time slot. Nevertheless, the presence of sensing and prediction errors can result in the misalignment of the beam with the actual target location. As a remedy, near-field beam split can enhance the robustness of the tracking algorithm by covering more potential locations in the next time slot and reducing the probability of beam misalignment.
\end{itemize}

\section{Conclusions and Future Research}
In this article, the impact of the unignorable near-field effect in the THz band on communications and sensing is investigated. The new characteristics caused by the near-field effect have been discussed, which introduce new opportunities and challenges for both THz communications and sensing. Specifically, apart from the ultra-large available bandwidth resources, THz communications can also exploit spherical wave propagation to enhance communication capacity. Furthermore, with the optics-like ultra resolution \cite{huawei2022} as well as the distance-aware ability in the large near-field region, THz sensing has the potential to provide high-fidelity three-dimensional (3D) reconstructions of the environment for smart applications.
However, it is crucial to note that research on THz near-field communications and sensing is still in its nascent phase. To illustrate, several existing open research issues can be exemplified as follows.

\begin{itemize}
    \item \textbf{THz Near-Filed Integrated Sensing and Communications:}
    Conventionally, communications and sensing are carried out by exploiting orthogonal resource blocks. Recently, ISAC, wherein both functions can utilize the same resource blocks and hardware platforms, has gained significant attention due to its potential for integration and coordination gain \cite{huawei2022}. Although the near-field ISAC has been initially investigated in \cite{wang2023near}, numerous challenges remain unresolved. One such challenge is the contradictory impact of the near-field beam split effect on communication and sensing. Effectively addressing the coordination of this beam split effect within the THz near-field ISAC system remains an open and unanswered question.

    \item \textbf{RIS/STAR-RIS aided THz Near-Field Communications and Sensing:}
    As discussed in the previous sections, both THz near-field communications and sensing highly rely on the LoS channels. However, the situation is not as expected in many cases, as the LoS channels frequently encounter obstructions. To address this challenge, reconfigurable intelligent surfaces (RISs) offer a viable solution by enabling the establishment of a virtual LoS channel through signal reflection for both communications and sensing. As a further advance, the concept of simultaneously transmitting and reflecting (STAR)-RIS has been proposed, allowing for full-space coverage and facilitating far-field communications and sensing \cite{wang2023stars}. This inspires research interest in RIS/STAR-RIS aided THz near-field communications and sensing.

    \item \textbf{Holographic THz Near-Field Communications and Sensing:}
    With the development of metamaterials, the concept of holographic antennas has emerged. These antennas employ a significantly reduced antenna spacing, which can be considerably shorter than half the wavelength, allowing for the approximation of the antenna array as continuous. Consequently, holographic antennas exhibit a more significant near-field effect compared to conventional antennas, which can benefit the application of THz near-field communications and sensing. For instance, holographic antennas have the capability to amplify the near-field multiplexing gain in THz MIMO communications and enhance the resolution of near-field sensing owing to the large number of antenna elements.
    
\end{itemize}

\bibliographystyle{IEEEtran}
\bibliography{reference/mybib}

\end{document}